\documentclass[%
 reprint,
 amsmath,amssymb,
 aps,
]{revtex4-1}

\usepackage{graphicx}
\usepackage{dcolumn}
\usepackage{bm}
\usepackage{xcolor}
\usepackage{hyperref}

\usepackage{amssymb, amsmath}
\usepackage{dutchcal}
\usepackage{braket}
\usepackage{calrsfs}
\DeclareMathAlphabet{\pazocal}{OMS}{zplm}{m}{n}
\usepackage{xfrac}    

\newcommand{\nm}{\ensuremath{N_{\mathrm{max}}}}
\newcommand{\w}{\ensuremath{\hbar\omega}}
\newcommand{\YN}{\ensuremath{Y\!N}}
\newcommand{\NN}{\ensuremath{N\!N}}
\newcommand{\mathleft}{\@fleqntrue\@mathmargin0pt}
\newcommand{\Lnn}{\ensuremath{\Lambda nn}}
\newcommand{\LamNN}{\ensuremath{\Lambda_{\NN}}}
\newcommand{\LamYN}{\ensuremath{\Lambda_{\YN}}}
\newcommand{\nnlosim}{NNLO\textsubscript{sim}}
\newcommand{\Tmax}{\ensuremath{T_{\rm Lab}^{\rm max}}}
\newcommand{\Tlab}{\ensuremath{T_{\rm lab}}}

\begin{document}

\preprint{APS/123-QED}

\title{Study on $\Lambda nn$ Bound State and Resonance}

\author{Thiri Yadanar Htun$^{1, 2}$}
\email[]{thiriyadanarhtun@gmail.com}
\author{Yupeng Yan$^{1}$}
\email[]{yupeng@g.sut.ac.th}

\affiliation{$^1$School of Physics and Center of Excellence in High Energy Physics and Astrophysics, Suranaree University of Technology, Nakhon Ratchasima 30000, Thailand\\
$^2$Department of Physics, University of Mandalay, 05032 Mandalay, Myanmar}

\date{\today}

\def\smath#1{\text{\scalebox{.9}{$#1$}}}
\def\sfrac#1#2{\smath{\frac{#1}{#2}}}

\begin{abstract}
We perform the ab initio no-core shell model (NCSM) calculation to investigate the bound state problem of three-body $\Lambda nn$ system in chiral next-to-next-to-leading-order NN and chiral leading-order YN interactions.  The calculations show that no $\Lambda nn$ bound state exists,  but predict a low-lying $\Lambda nn$ resonant state near the threshold with the energy of $E_r= 0.124$ MeV and the width of about $\Gamma=1.161$ MeV.  In searching for $\Lambda nn$ resonances, we extend the NCSM calculation to the continuum state by employing the J-matrix formalism in the scattering theory with the hyperspherical oscillator basis.

\end{abstract}

\pacs{Valid PACS appear here}

\maketitle

\section{\label{sec:level1}INTRODUCTION}

\indent Microscopic calculations of few- and many-body systems with strangeness have been a focus in hypernuclear physics to explore the new dynamical features of the structure of hypernuclei and to improve understanding of hyperon-nucleon interactions. Indeed, hyperon-nucleon scattering data is very limited to fully determine the YN interactions.  The existing data of few body hypernuclei could provide the important constrain on YN interaction.  In hypernuclear physics, hypetriton is used as the first testing ground for YN interaction. It is the simplest and weakly bound hypernuclear system with $\Lambda$ binding energy about $\sim$ 0.13 MeV \cite{Davis:2005mb}.  It seems like a lambda bound to deuteron core in the study of the spin-triplet NN interaction \cite{Miyagawa:1995sf}.  In the \Lnn{} system, two neutrons interact in spin-singlet state and its strength is weaker than that in spin-triplet state.  The strength of $\Lambda n$ is also not sufficient to form a bound system.  It expects that the existence of a neutral bound state of two neutrons and a hyperon is improbable.  But instead, three-body \Lnn{} resonance may exist and that could be used to constrain the YN interaction.  If \Lnn{} system were a lightest neutron-rich bound system,  it would provide significant information of $\Lambda n$ interaction and a better understanding of the nature of $\Lambda$N-$\Sigma N$ coupling.

\indent There have been a number of theoretical calculations for the \Lnn{} system as a serious doubtful bound state problem.  Nonexistence of \Lnn{} bound state was first revealed by Dalitz and Downs \cite{Downs:1959zz} using a variational approach.  Garcilazo \cite{Garcilazo_1987} investigated the \Lnn{} system by solving Faddeev equations using YN and NN interactions derived from a chiral constituent quark model and revealed that \Lnn{} bound system was not found. Later the various approach such as hyperspherical harmonics (HH)\cite{BELYAEV2008210},  Faddeev calculations \cite{Miyagawa:1995sf,Afnan:2015ahc,Filikhin:2016raj,Garcilazo:2007ss,Garcilazo:2014lva,Kamada2016,Gibson:2019occ},  variational calculations \cite{Hiyama:2014cua},  pionless effective field theory \cite{Ando:2015fsa,Hildenbrand:2019sgp,Schafer:2020jnb} with various kinds of baryon-baryon interactions have been used to analyze the \Lnn{} system and all reported that it is highly unlikely to form a bound system in the theoretical analysis without a significant altering nuclear and hypernuclear forces.

\indent The $_\Lambda^3n$ hypernucleus could not be produced in the earlier experiments due to no charge of its bound state.  However,  the HypHI collaboration at GSI \cite{Rappold:2013jta} reported the first evidence of the existence of the $_\Lambda^3n$ bound state from analysis of the observed two- and three-body decays mode without describing any statement for the value of binding energy. Their observation was inconsistent with the claim of the above theoretical analysis.

\indent In this paper,  we analyze the \Lnn{} bound state problem using the ab initio no-core shell model (NCSM) \cite{Barrett:2013nh, Wirth:2014apa, Wirth:2017bpw} technique.  The calculation of the \Lnn{} system $(J^\pi =1/2^+ ,T=1)$ is performed in Jacobi coordinate HO basis using the NN and YN interactions derived from chiral effective field model.  In the extension into the continuum state, we apply the SS-HORSE \cite{Mazur:2015djl,Shirokov:2016thl,Blokhintsev:2017rae,Lurie:2003kj,Shirokov:2018nlj,Mazur:2019ots} formalism, which is a single state harmonic oscillator representation of scattering equations, to calculate the low-energy phase shifts and scattering amplitudes at the NCSM eigenenergies by employing hyperspherical harmonic oscillator basis.  The low-lying \Lnn{} resonance energy and width are extracted from the scattering amplitude parametrization.  The NCSM-SS-HORSE method \cite{Shirokov:2016ywq} has been successfully applied to study a tetraneutron unbound system considered as true four-body scatterings. Here we first apply this method to study the three-body system with strangeness.

\section{\label{sec:level2}NCSM-SS-HORSE Formalism}

\indent The hypernuclear Hamiltonian for two nucleon and a hyperon system can be written
\begin{equation}
  \label{eq:h}
  H = -\sum_{i=1}^3\frac{\hbar^2}{2m_i}\vec{\nabla}^2_i +
  V_{NN}(\vec{r}_1,\vec{r}_2) + \sum_{i=1}^2 V_{YN{}}(\vec{r}_i,
  \vec{r}_3) + \Delta M,
\end{equation}
where the coordinates $\vec{r}_i$ and masses $m_i$ are for the two nucleon with $i=1,  2$ and the hyperon with $i=3$.
We work with nonrelativistic two-body \NN{} and \YN{} potentials,  employing the leading-order chiral hyperon-nucleon  interactions with regulator cutoff \LamYN{} = 600 MeV \cite{Polinder:2006zh} and a family of 42 different nuclear interactions at next-to-next-to-leading order (also called chiral \nnlosim{} family of NN interactions) \cite{Carlsson:2015vda}. These nuclear interactions were constructed by varying the
chiral regulator cutoff \LamNN{} between 450 and 600 MeV in steps of
25 MeV and the truncation of the input \NN{} scattering $\Tlab \leq \Tmax$ between 125 and 290 MeV in six steps,   which were obtained from a simultaneous optimization of all 26 low-energy constants (LECs) to different sets of NN and $\pi$N scattering plus bound state observables \cite{Carlsson:2015vda}.  In this work,  we mainly use the \NN{} interactions with \LamNN{}=500 and \Tmax{}=290 MeV. The effect of $\Lambda N$-$\Sigma N$ coupling is taken into account \cite{Wirth:2017bpw}.

\indent In NCSM,  three active particles are considered in the three-dimensional harmonic oscillator (HO) basis.  In the construction of HO basis states for such a few-body \Lnn{} system,  it is more effective to use the relative Jacobi coordinates where the center of mass (c.m.) coordinate $\vec{\xi}_{0} $ is separated, which allows us to perform NCSM calculations in large model space.
The relative Jacobi coordinates in terms of the rescaled version of the single-particle coordinates $\vec{x}_i = \sqrt{m_i} \vec{r}_i$ are defined as
\begin{equation}
  \label{eq:xi}
  \begin{split}
    \vec{\xi}_{1} &=\sqrt{\sfrac{1}{2}}\left(\vec{x}_1 - \vec{x}_2\right),\\
    \vec{\xi}_{2} &=\sqrt{\sfrac{2m_{N} m_Y}{2m_{N}+m_Y}}\left[\sfrac{1}{2\sqrt{m_{N}}}(\vec{x}_1+\vec{x}_2)-\sfrac{1}{\sqrt{m_Y}}\vec{x}_3\right],\\
  \end{split}
\end{equation}
where $m_N$ and $m_Y$ are the masses of nucleon and hyperon.   $\vec{\xi}_1$ is the relative coordinate of the two-nucleon pair and $\vec{\xi}_2$ is the relative coordinate of the hyperon with respect to the c.m.  of the two-nucleon pair.
Following the general Jacobi coordinate formulation in Ref. \cite{Wirth:2017bpw},  we construct the JT-coupled HO basis states for the system of a two-nucleon pair and a hyperon,
\begin{equation}\label{basis_nn}
\ket{(n_{NN}(l_{NN}s_{NN})j_{NN}t_{NN} ,\pazocal{N}_Y \mathcal{L}_Y  \pazocal{J}_Y \pazocal{T}_Y)JT},
\end{equation}
depending on the coordinates $\vec{\xi}_{1}$ and $\vec{\xi}_{2}$ respectively.  $ n_{NN},\;l_{NN},\;s_{NN},\;j_{NN},\;t_{NN}$ $ (\pazocal{N}_Y,\;\mathcal{L}_Y,\; \pazocal{J}_Y,\;  \pazocal{T}_Y)$ are the HO radial quantum number,  orbital angular momentum,  spin,  angular momentum and isospin of the relative two-nucleon (hyperon) state.  $J$ and $T$ are the total angular momentum and total isospin of the system.  The basis \eqref{basis_nn} is antisymmetrized with respect to the exchange of two nucleon by restricting the two nucleon relative quantum numbers with the condition $(-1)^{l_{NN}+s_{NN}+t_{NN}} = -1$.
The basis (\ref{basis_nn}) is suitable for evaluating two-body \NN{} interaction matrix elements but not for evaluating two-body \YN{} interaction matrix elements.

\indent For a subsystem including \YN{} pair and a nucleon,  another set of Jacobi coordinate is correspondingly introduced,
\begin{equation}
  \label{eq:eta}
  \begin{split}
    \vec{\eta}_{1} &=\sqrt{\sfrac{(m_{N}+m_Y)m_{N}}{2m_N+m_Y}}
      \bigg[\sfrac{1}{\sqrt{m_{N}}} \vec{x}_1\\
      &-
      \sfrac{1}{(m_{N}+m_Y)}(\sqrt{m_{N}}\vec{x}_2+\sqrt{m_Y}\vec{x}_3)\bigg],\\
    \vec{\eta}_{2} &=\sqrt{\sfrac{m_{N} m_Y}{m_{N}+ m_Y}}\left(\sfrac{1}{\sqrt{m_{N}}}\vec{x}_2 -\sfrac{1}{\sqrt{m_Y}}\vec{x}_3\right),
    \end{split}
\end{equation}
where $\vec{\eta}_{1}$ is the relative coordinate of a nucleon with respect to the c.m. of the \YN{} pair and $\vec{\eta}_{2}$ is the relative coordinate of the \YN{} pair.
By using orthogonal transformation,  the antisymmetrized HO basis  (\ref{basis_nn}) can be expanded as
\begin{equation} \label{basis12}
	\begin{split}
	&\ket{(n_{NN}(l_{NN}s_{NN})j_{NN}t_{NN},\pazocal{N}_Y \mathcal{L}_Y \pazocal{J}_Y\pazocal{T}_Y)JT}\\
 	&=\sum_{LS}\hat{L}^2\hat{S}^2 \hat{j}_{NY}\hat{\pazocal{J}}_{N}\hat{j}_{NN}\hat{\pazocal{J}}_Y (-1)^{s_{NY}+\sfrac{1}{2}+s_{NN}+\sfrac{1}{2}+\mathcal{L}_{N}+\mathcal{L}_Y}\\
 	&\times
\begin{Bmatrix}
l_{NY} & s_{NY} & j_{NY}\\
\mathcal{L}_{N} & \sfrac{1}{2} & \pazocal{J}_{N}\\
L & S & J
\end{Bmatrix}
\begin{Bmatrix}
l_{NN}                   & s_{NN}         & j_{NN}\\
\mathcal{L}_Y  & \sfrac{1}{2} & \pazocal{J}_Y\\
L & S & J
\end{Bmatrix}
\begin{Bmatrix}
\sfrac{1}{2}  & \sfrac{1}{2} & s_{NN}\\
\sfrac{1}{2}  &  S                   & s_{NY}
\end{Bmatrix}   \\
	&\times(-1)^{t_{NY}+\pazocal{T}_N+t_{NN}+\pazocal{T}_Y} \hat{t}_{NY}\hat{t}_{NN}
\begin{Bmatrix}
\sfrac{1}{2}       &  \sfrac{1}{2}  & t_{NN}\\
\pazocal{J}_Y  &    T                   & t_{NY}
\end{Bmatrix}\\
	&\times \braket{n_{NY}l_{NY}\pazocal{N}_N\mathcal{L}_N|n_{NN}l_{NN}\pazocal{N}_Y \mathcal{L}_Y}_{d=\frac{2m_N+m_Y}{m_Y}}\\
&\times \ket{(n_{NY}(l_{NY}s_{NY})j_{NY}t_{NY},\pazocal{N}_N\mathcal{L}_N \pazocal{J}_N)JT},
	\end{split}
\end{equation}
in terms of HO basis states
\begin{equation}\label{basis2}
\ket{(n_{NY}(l_{NY}s_{NY})j_{NY}t_{NY},\pazocal{N}_{N} \mathcal{L}_{N} \pazocal{J}_{N} )JT},
\end{equation}
depending on the coordinates $\vec{\eta}_2$ and $\vec{\eta}_1$ respectively.
The general HO bracket $\braket{n_{NY}l_{NY}\pazocal{N}_{N}\mathcal{L}_{N} \vert n_{NN}l_{NN}\pazocal{N}_Y \mathcal{L}_Y}_d$ follows the agreement of Ref. \cite{Kamuntavicius:2001pf}.  $YN$ interaction matrix elements involving  $\Lambda$ and $\Sigma$ hyperons are evaluated in the antisymmmetrized basis~\eqref{basis_nn} through its expansion in the basis~\eqref{basis2} as
\begin{equation}
  \label{eq:vny}
  \braket{\sum_{i=1}^2 V_{YN}(\vec{r}_i,
  \vec{r}_3)} = 2 \braket{V_{YN}(\vec{\eta}_2)},
\end{equation}
where the matrix elements on the right-hand side are diagonal in all quantum numbers of the basis states~\eqref{basis2},  except for $n_{NY}$ and $l_{NY}$.  The lowest eigenstates of the \Lnn{} system are calculated by the diagonalization of the truncated Hamiltonian matrix.

\indent To look for resonances, we extend our study to the continuum state by employing J-Matrix formalism, also known as Harmonic oscillator representation of scattering equation (HORSE) formalism, which arms one to study continuum spectrum using only positive energies obtained from bound state approach like NCSM applying HO basis. The HORSE method can be used to describe the open channels in the external subspace while the internal subspace is associated with the NCSM approach.  For details of the HORSE formalism, 
we may refer to Refs. \cite{Blokhintsev:2017rae,BANG2000299}. 

\indent In the extension into continuum, the three-body extension of the J-matrix formalism for all three-body decay channels is very complicated.  We apply the democratic decay approximation (also known as true three-body scattering or $3\rightarrow3$ scattering) \cite{Zaitsev1998TrueMS} which employs the hyperspherical harmonic (HH) basis to describe the \Lnn{} system decaying through only three-body break-up channel
and it does not allow for other possible two-body channels associated with two-body sub-bound states.

\indent The hyperspherical oscillator basis can be labeled as $\ket{\kappa K \gamma}$, where $\kappa$ is the principal quantum number and $K$ is the hypermomentum,  $\gamma \equiv \{l_1,  l_2,  L , s_1,  s_2,  S,  t_1,  t_2,  T \}$ collects all possible quantum numbers corresponding to the Jacobi coordinates for a three-body system.
The external subspace is spanned by hyperspherical oscillator functions with $N \equiv 2\kappa+K > \nm$ where the Hamiltonian $H=T$ is used.   Here $\nm$ is the maximum number of excitation quanta defining the many-body NCSM basis space.
Because of high centrifugal barrier $\pazocal{L}(\pazocal{L}+1)/\rho^2$,  the HH states with larger K can be neglected in the case of no sub-bound \Lnn{} system [$\rho$ is hyper radius with the mass scaled Jacobi coordinates and $\pazocal{L}=K+3/2$ is the effective momentum].
It is adequate to consider a single hyperspherical channel with minimum hypermomentum $K_{min}=0$ to describe democratic three-body decays.  

\indent We follow the SS-HORSE approach \cite{Blokhintsev:2017rae,Shirokov:2016thl,Shirokov:2016ywq} to compute the $3\rightarrow3$ scattering phase shifts at the eigenenergies $E_\nu> 0$ obtained directly from NCSM calculation,
\begin{equation}\label{eq:Ps}
\tan\,\delta(E_\nu)=-\frac{S_{\nm+2,\pazocal{L}}(E_\nu)}{C_{\nm+2,\pazocal{L}}(E_\nu)},
\end{equation}
where $S_{N\pazocal{L}}$ and $C_{N\pazocal{L}}$ are regular and irregular solutions of free Schr\"{o}dinger equation in the hyperspherical oscillator representation, which can be applied in the case of arbitrary $\pazocal{L}$ (both integer and half integer),
taking simple analytical expressions \cite{Zaitsev1998TrueMS,Lurie:2003kj,Shirokov:2016thl}
\begin{equation}\label{eq:Snl}
\begin{split}
 S_{N \pazocal{L}}(E)=& \sqrt{\frac{(N-\pazocal{L}+\dfrac{3}{2})!}{\lambda~ \Gamma (\dfrac{N}{2}+\dfrac{\pazocal{L}}{2}+\dfrac{9}{4})}} ~q^{\pazocal{L}+1} ~e^{-\frac{q^2}{2}}\\
 & L^{\pazocal{L}+\frac{1}{2}}_{(N-\pazocal{L}+\frac{3}{2})/2} (q^2),\\
\end{split}
\end{equation}
\begin{equation} \label{eq:Cpm_nl}
	\begin{split}
    C^{(\pm)}_{N \pazocal{L}}(E)= & \frac{1}{\pi \sqrt{\lambda}} \sqrt{(N-\pazocal{L}+\dfrac{3}{2})!~ \Gamma(\dfrac{N}{2}+\dfrac{\pazocal{L}}{2}+\dfrac{9}{4})}\\
   & \Psi(\dfrac{N}{2}+\dfrac{\pazocal{L}}{2}+\dfrac{9}{4},\pazocal{L}+\frac{3}{2};e^{\mp i\pi}q^2) \\
   & q^{\pazocal{L}+1} ~e^{\frac{q^2}{2}}~e^{\mp i\pi (\pazocal{L}+\frac{1}{2})},
	\end{split}
\end{equation}
\begin{equation} \label{eq:Cnl}
C_{N \pazocal{L}}(E)=\dfrac{1}{2}\Big(C_{N \pazocal{L}}^{(+)}(E)+C_{N \pazocal{L}}^{(-)}(E)  \Big),
\end{equation}
\begin{figure}[h!]
	\centering
	\includegraphics[width=1\linewidth]{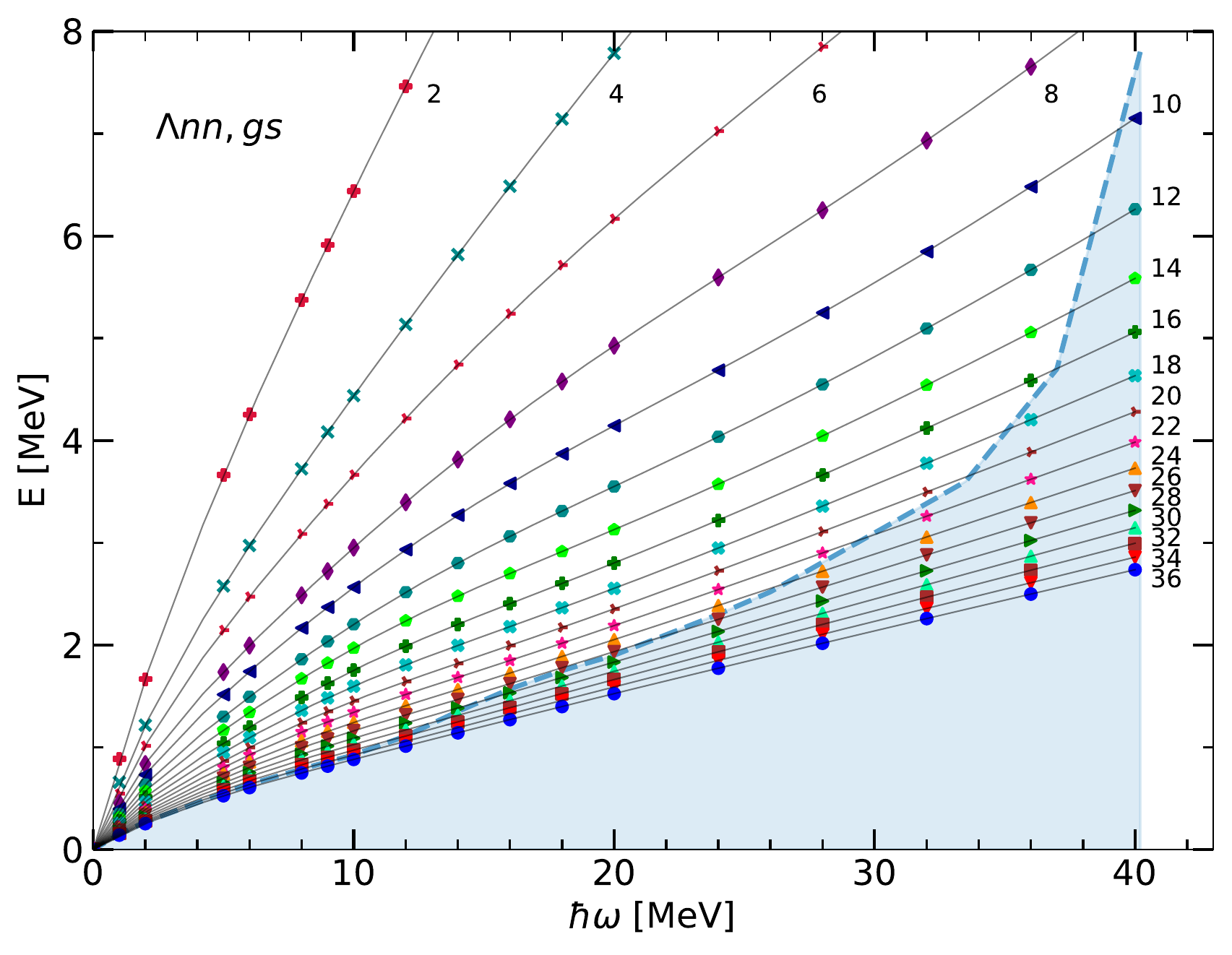}
	\caption{The eigenenergies of the NCSM Hamiltonian with various model space sizes \nm{} as a function of oscillator frequency \w. The numbers at the end of each line represent \nm.  The blue shaded area shows the selected energies for parametrization of the scattering amplitude.}
	\label{fig:Econvergence}
\end{figure}
\begin{figure}[h!]
    \centering
	\includegraphics[width=1\linewidth]{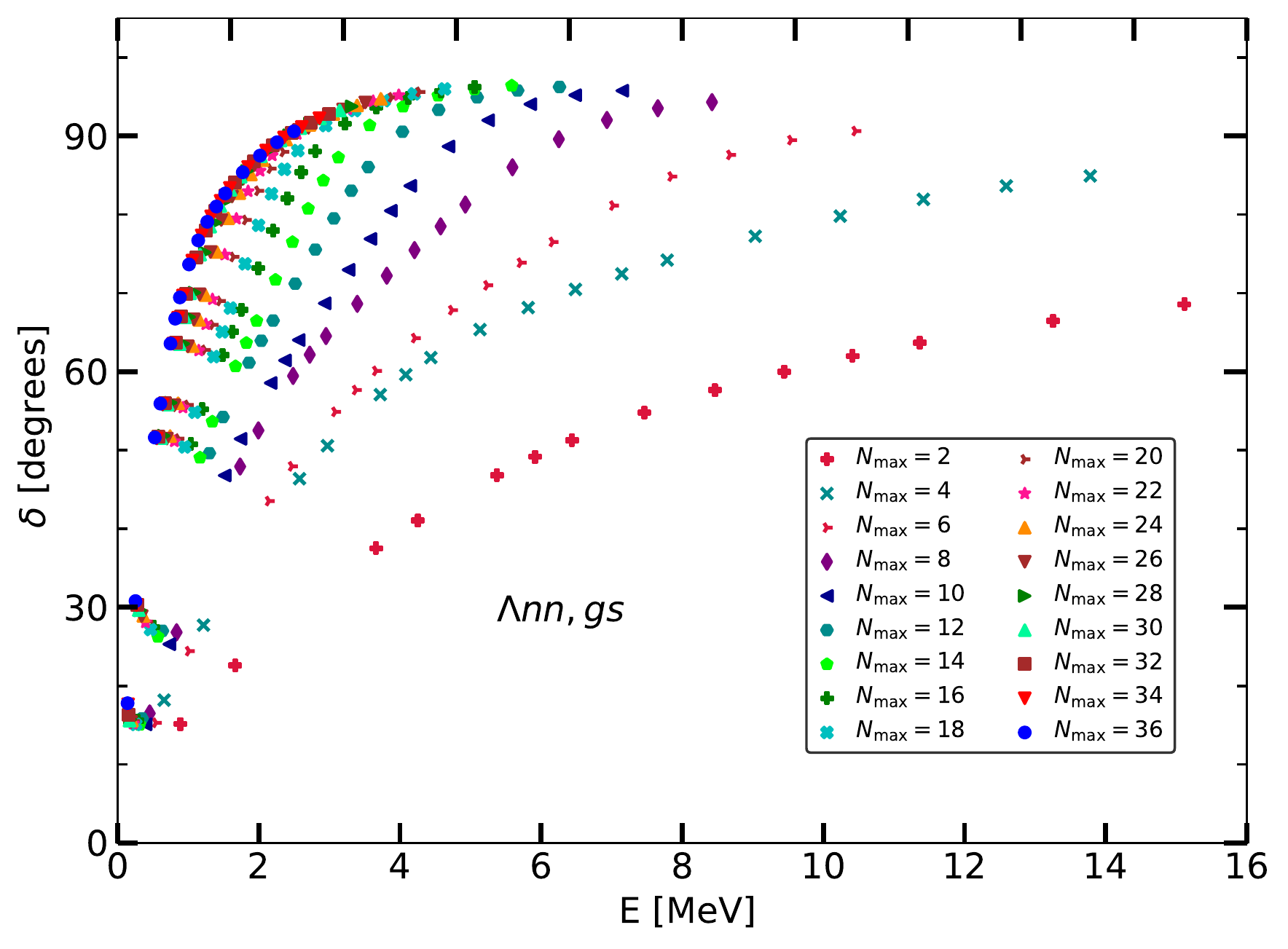}
	\setlength{\belowcaptionskip}{-10pt}
	\caption{$3\rightarrow3$ scattering phase shifts obtained directly from the NCSM eigenstates using Eq. \eqref{eq:Ps}.}
	\label{fig:PS}
\end{figure}

\noindent where $q=\sqrt{\frac{2E}{\w}}$ is dimensionless momentum,  $L^{\pazocal{L}+\frac{1}{2}}_{\kappa}(x)$ is the associated Laguerre polynomial,  $\lambda=\sqrt{\frac{m\omega}{\hbar}}$ is the oscillator radius at $\Psi(a,c;x)$ which is the Tricomi function.

The SS-HORSE scattering amplitude for neutral particles may be calculated in the standard way,
\begin{equation}\label{eq:f(E)}
f(E_\nu) q =\frac{1}{(\rm{cot}\,\delta_{\pazocal{L}}(E_\nu)-i)}.
\end{equation}
We parameterize the scattering amplitude in the method proposed in \cite{Cho:1993wh} for the case that a resonance is not sharp, but both the potential scattering (non-resonant background) and resonance contribution are not negligible. The scattering amplitude may be parametrized as
\begin{equation}\label{eq:F(E)_function}
  \begin{aligned}
	F(E)q & = e^{i\delta_0(E)}\sin\delta_0(E)+\frac{-\Gamma/2}{E-E_r+i\Gamma/2} e^{2i\delta_0(E)},
  \end{aligned}
\end{equation}
where  $\delta_0(E)$ is the potential scattering phase shift, depending on the energy $E$.  We will fit the SS-HORSE scattering amplitude by the complex-valued function $F(E)q$ in the next section to determine the form of the $\delta_0(E)$ and derive the resonance energy $E_r$ and width $\Gamma$.

\section{\label{sec:level3} Results and Discussion }

\indent The \Lnn{} system is analyzed using the NCSM approach with chiral \nnlosim{} NN and LO YN interactions. The NCSM computational model space is characterized by a chosen maximal total HO quanta $\nm^{tot}$, that is,
\begin{equation}
2n_{NN}+l_{NN}+2\pazocal{N}_{Y} + \pazocal{L}_{Y} \leq \nm^{tot} \equiv \nm+N_0,
\end{equation}
where the minimal possible number of HO quanta is $N_0=0$.  In \Lnn{} case,  $\nm^{tot}=\nm$.
We have computed the total energy of \Lnn{} system in the oscillator basis with model space truncations $\nm \leq 36$,  and in the range of the HO frequencies 1 MeV $\leq\hbar\omega\leq 40$ MeV.  It is found that there is no \Lnn{} bound system.
The \Lnn{} ground-state energy as a function of the model space truncation \nm{} and HO frequency \w{} is presented in Figure~\ref{fig:Econvergence}.
The NCSM energies decrease with increasing \nm{} and with decreasing \w{}.
Our model used here can reproduce well the binding energy of hypertriton \cite{Htun:2021jnu} and also for s-shell hypernuclei, $_\Lambda^4 H$ and $_\Lambda^4 He$, which will be a future publication.

\begin{figure}[h!]
    \centering
	\includegraphics[width=1\linewidth]{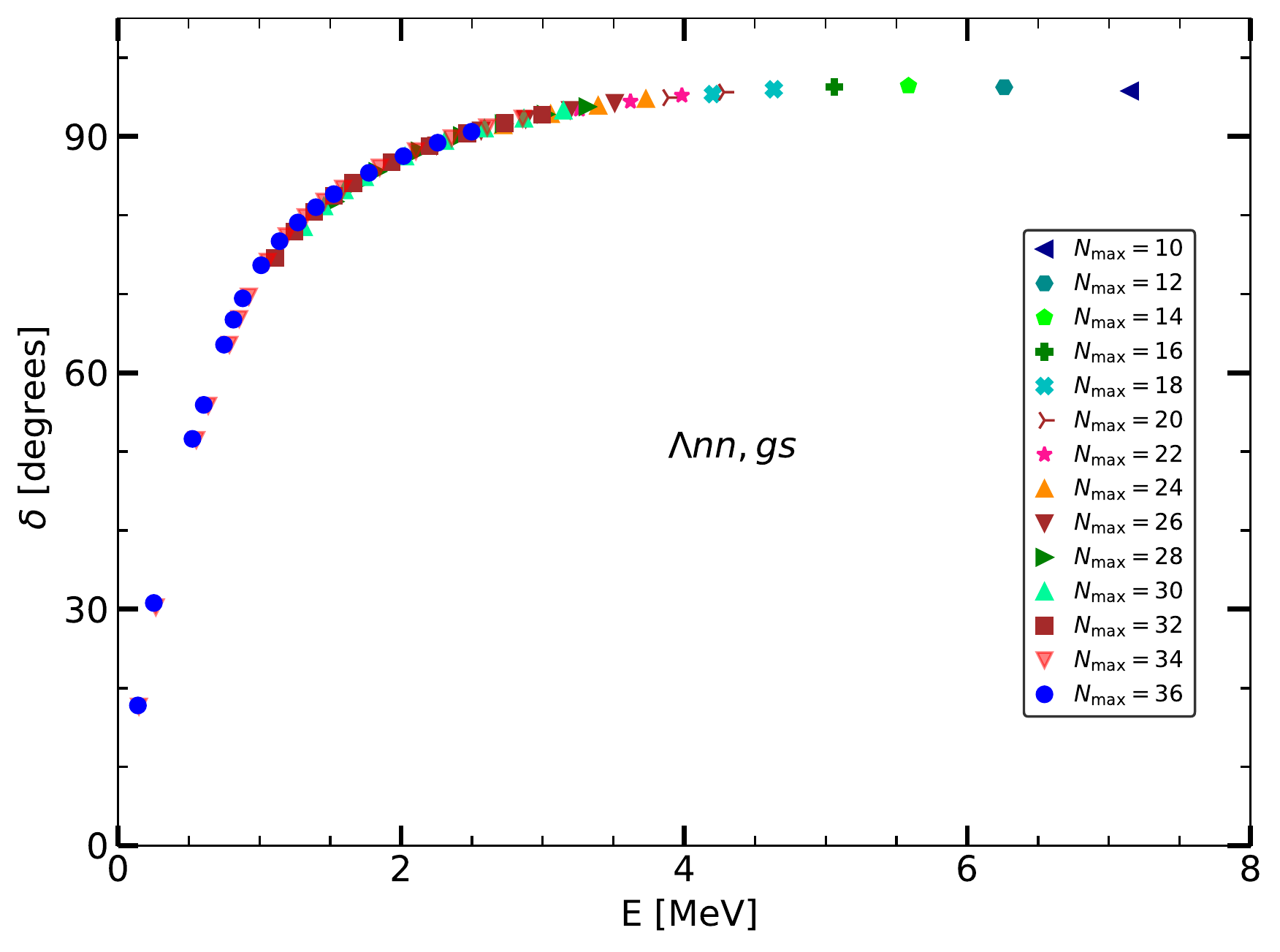}
	\caption{$3\rightarrow3$ scattering phase shifts obtained from selected NCSM eigenstates with $\nm \in [10,36]$ for scattering amplitude parametrization.}
	\label{fig:PS-converge}
\end{figure}

\indent The SS-HORSE phase shifts covering all computed NCSM energies calculated by using Eq. ~\eqref{eq:Ps} are shown in Figure~\ref{fig:PS}.  The phase shifts obtained with smaller $\nm$ lie in a wide energy region as the obtained \Lnn{} ground-state energies spread widely.  With \nm{} increasing,  however,  the obtained $\Lambda nn$ ground-state energies converge to lower values, as shown in Figure~\ref{fig:Econvergence},  and hence the corresponding phase shifts shift to the lower energy region. The first convergence of phase shifts is achieved at smaller energies with larger \nm{},  almost the same results at \nm{} = 34 and 36 MeV.
We follow the selection procedure of Ref. \cite{Shirokov:2016ywq,Shirokov:2016thl,Mazur:2016gte} and select a set of eigenvalues $E_\nu$ from the \nm= 10-36 model spaces,  which is illustrated by the shaded area in Figure~\ref{fig:Econvergence},  to produce a single smooth curve of phase shifts for parametrization.
 The SS-HORSE phase shifts corresponding to these selected smaller eigenvalues are plotted in Figure~\ref{fig:PS-converge}.

\begin{figure}[h!]
    \centering
	\includegraphics[width=1\linewidth]{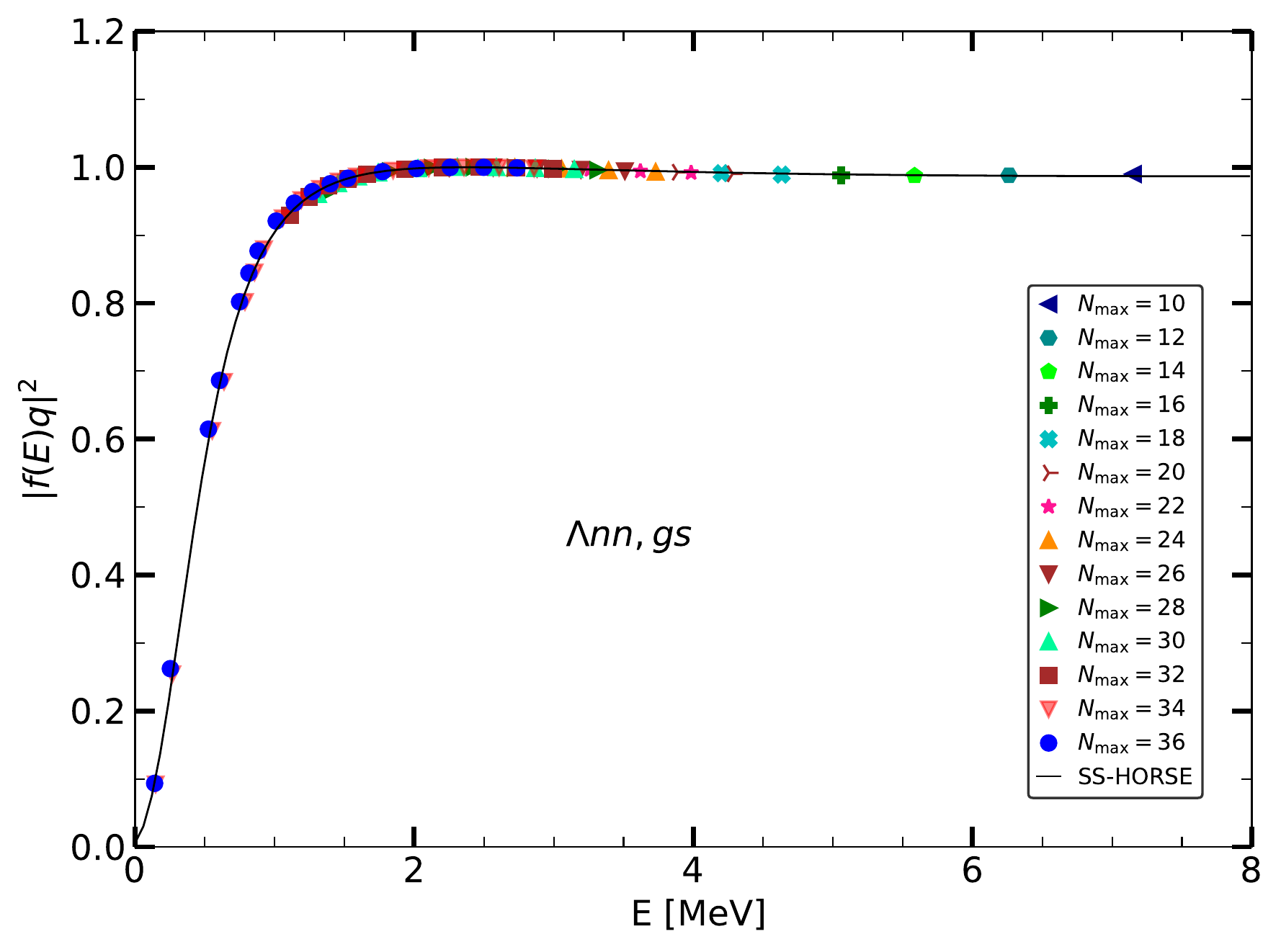}
	\setlength{\belowcaptionskip}{-10pt}
	\caption{The scattering amplitude $|f(E)q|^2$ using Eq. \eqref{eq:f(E)} obtained from NCSM eigenstates (symbol). The solid line shows the parametrization of scattering amplitude in Eq. ~\eqref{eq:F(E)_function}.}
	\label{fig:ScatteringAmplitude}
\end{figure}

\indent We compute the SS-HORSE low-energy scattering amplitude for the purpose of extracting the resonance parameters from scattering amplitude parametrization. The function $|f(E_\nu )q|^2$ of the scattering amplitude given in Eq.  \eqref{eq:f(E)} is shown by symbols in Figure~\ref{fig:ScatteringAmplitude}. The fitting to the SS-HORSE result $|f(E_\nu )q|^2$ by the function $|F(E)q|^2$ leads the $\delta_0(E)$ to the form
\begin{equation}
\delta_0(E)=a_0+a_2(\sqrt{E})^{2}+a_4(\sqrt{E})^{4},
\end{equation}
with the adjustable parameters $a_0=1.856$,  $a_2=-0.014$ MeV${}^{-1}$,  $a_4=2.959 \times 10^{-4}$ MeV${}^{-2}$.  The resonance energy and width are derived,   $E_r= 0.124$ MeV and $\Gamma=1.161$ MeV.  The result is in good agreement with those in Ref.  \cite{Filikhin:2016raj,Gibson:2019occ} and lies within the estimated range of the location and width of a \Lnn{} pole \cite{Schafer:2021mit}. We look forward to the results of \Lnn{} bound and resonance states from the ongoing experiment (E12-17-003) at Jefferson Lab (JLab) \cite{Tang:2018}. Such \Lnn{} bound and resonance states, if any, are expected to provide new perspective on $\Lambda n$ interactions.

\section*{Summary}
We have performed ab initio no-core shell model calculations for the \Lnn{} system $(J^\pi =1/2^+ ,T=1)$ without tuning the strength of realistic NN and YN potentials at various \nm{} and \w{} values with full inclusion of $\Lambda N$-$\Sigma N$ coupling, and found that no bound state exists. To look for resonance states of the \Lnn{},  we have applied the NCSM-SS-HORSE technique to calculate the \Lnn{} scattering phase shifts which suggest a \Lnn{} resonant state at energy $E_r= 0.124$ MeV and $\Gamma=1.161$ MeV.  Further theoretical studies and experimental searches for \Lnn{} resonances would be of great benefit of constraining $\Lambda n$ interactions.

\begin{acknowledgments}
We are grateful to Daniel Gazda for providing us with the NCSM codes used to compute the \Lnn{} eigenenergies.
 This work is supported by the Royal Golden Jubilee Ph.D. Program jointly sponsored by Thailand International Development Cooperation Agency,  Thailand Research Fund, Swedish International Development Cooperation Agency and International Science Programme (ISP) at Uppsala University under Contract No. PHD/0068/2558.  
 
\end{acknowledgments}

\nocite{*}

\bibliography{bibl}

\end{document}